\documentclass{an}
\usepackage{graphicx}
\usepackage{times}
\usepackage{fancyhdr}
\def\lesssim{\mathrel{\hbox{\rlap{\hbox{\lower2pt\hbox{$\sim$}}}\raise2pt\hbox{$<$}}}}
\def\grtsim{\mathrel{\hbox{\rlap{\hbox{\lower2pt\hbox{$\sim$}}}\raise2pt\hbox{$>$}}}}

\begin{document}

\title{The TOOT00 Redshift Survey of Radio Sources}

\author{
Eleni Vardoulaki\inst{1}, Steve Rawlings\inst{1}, Gary J. Hill\inst{2}, Steve 
Croft\inst{3}, Kate Brand\inst{4}, Julia Riley\inst{5}, Chris Willott\inst{6}
}
\institute{ 
Astrophysics, Denys Wilkinson Building, Keble Road, Oxford, OX1 3RH, UK
\and
University of Texas at Austin, 1 University Station, C1402, Austin, TX
78712, USA
\and
Institute of Geophysics and Planetary Physics, Lawrence Livermore National 
Laboratory L-413, 7000, East Avenue, Livermore, CA 94550
\and
National Optical Astronomy Observatory, Tuscon, AZ 85726-6732
\and
Cavendish Astrophysics, Department of Physics, Madingley Road, Cambridge, 
CB3 OHE
\and
Herzberg Institute of Astrophysics, National Research Council, 5071 West 
Saanich RD, Victoria, BC V9E 2E7, Canada
}

\date{Received; accepted; published online}

\abstract{ We present first results from the study of the TOOT00 region 
consisting of 47 radio sources brighter than 100 mJy at 151 MHz. We have 
81\% spectroscopic redshift 
completeness. From the $K-z$ diagram we deduce that the host galaxies are 
similar to $\sim$3 $L^{*}$ passively evolved elliptical galaxies and thus 
estimate the redshifts of the 9 sources without a 
secure spectroscopic redshift yielding a median redshift of 1.287. 
Above the RLF break we have a quasar fraction 
$f_{\rm q} \sim$ 0.3 although the quasars appear reddened; below the 
RLF break $f_{\rm q} \rightarrow$ 0 if we exclude flat-spectrum radio sources. 
We present a histogram of the 
number of TOOT00 radio sources versus their 
redshift which looks broadly 
like the Willott et al.\ (2001) prediction for TOOT, although the observed 
ratio of high to low redshift objects is somewhat lower than the 
prediction. }

\correspondence{eleniv@astro.ox.ac.uk, sr@astro.ox.ac.uk}

\maketitle

\section{Introduction}
In search of a better way to determine the evolution of the radio-source 
population and to study the Radio Luminosity Function (RLF) many researchers 
have defined different redshift surveys using various 
samples at decreasing flux density limits (for example, the 3CRR, 6CE and 7CRS 
surveys). The first comprehensive study of the RLF was carried out by Dunlop 
and Peacock (1990) who used several complete samples selected at 2.7 GHz with 
flux density limits between 2.0 and 0.1 Jy. For samples selected at low 
frequency the benchmark bright survey is the 3CRR survey 
(Laing, Riley \& Longair 1983). Various fainter surveys have followed, 
but the most complete in spectroscopic redshifts so far are the 6CE 
(Rawlings et al.\ 2001) and 7CRS samples (Willott et al.\ 2002).

The first demonstration of the need for a complete sample with a lower flux 
density limit than the 3CRR, 6CE and 7CRS samples was made by Willott et al.\ 
(2001), who argued that in order to determine the RLF at higher redshifts a 
new redshift survey based on a large ($\sim$1000 source) sample about five 
times fainter than the 7CRS was needed. This was the scientific basis for the 
TexOx-1000 (TOOT) survey (Hill \& Rawlings 2003), which aims to measure 
redshifts for $\sim$ 1000 sources from the 7C survey with a flux density limit 
of 100 mJy at 151 MHz. TOOT was designed to trace radio sources that lie at 
the break in the RLF at reasonably high redshift. 
This makes TOOT the first survey able to 
study the sources responsible for the bulk of the radio luminosity 
density of the universe at high redshift (Hill \& Rawlings 2003). The goal is 
to estimate the amount of energy injected in the intracluster medium (ICM) by 
radio sources and investigate how the radio-source population contributes to 
the entropy budget of the Universe (Rawlings 2003). The entropy budget of the 
Universe determines the 
structure and evolution of the intracluster medium (ICM) and records the 
thermodynamic history of the gas in clusters (Voit 2004). Furthermore, 
$\sim$10\% of the thermal energy in cluster baryons may originate from radio 
sources (e.g. Rawlings 2000); the study of radio sources by TOOT may help to 
explain the `excess' entropy inferred in the central regions of clusters 
(Ponman et al.\ 1999).

The TOOT survey will also probe the existence of a redshift cut-off (e.g. 
Willott et al.\ 2001) for high radio luminosity sources at $z >$ 2, and allow 
us to carry out studies of various radio source properties which have up to 
now been dogged by small number statistics. This is particularly important 
because TOOT probes radio sources near the RLF break which are typical radio 
sources in the same luminosity-weighted way that an $L^{*}$ galaxy is typical 
of normal galaxies. It is also important to investigate 
whether the low-luminosity population evolves at all (e.g. Clewley \& Jarvis 
2005). 

According to Faranoff \& Riley (1974) there are two classes of radio sources, 
FRI and FRII, where above 
$\log_{10}( L_{151{\rm MHz}}/{\rm W Hz}^{-1} {\rm sr}^{-1})$ = 25.5 lie the 
FRIIs and below that the FRIs. More recent classification agrees with a 
dual-population model where the less radio-luminous population is composed 
of FRIs and FRIIs with weak/absent emission lines, and the more radio-luminous 
population of strong-line FRII radio galaxies and quasars, where the division 
is at 
$\log_{10}( L_{151{\rm MHz}}/{\rm W Hz}^{-1} {\rm sr}^{-1})$ = 26.5 
(Willott et al.\ 2001) corresponding to the RLF break.

In this paper we present the scientific results derived from a preliminary 
study of one of the TOOT regions using optical, near-infrared and radio data. 
All the data will be presented elsewhere (Vardoulaki et al.\ in prep).

We use the convention for all spectral 
indices, $\alpha$, that flux density $S_{\nu} \propto \nu^{-\alpha}$,
where $\nu$ is the observing frequency. 
We assume throughout a low-density, $\Lambda$-dominated Universe in which
$H_{\circ}=70~ {\rm km~s^{-1}Mpc^{-1}}$, $\Omega_{\rm M}=0.3$ and $\Omega_
{\Lambda}=0.7$.

\section{The TOOT00 region}
The TOOT survey consists of several regions (Hill \& Rawlings 2003), but here 
we consider TOOT00 only, studying 47 radio sources with flux densities above 
100 mJy at 151 MHz. TOOT00 covers 1.42 deg$^{2}$ around RA = 00$^{\rm h}$ 
14$^{\rm m}$, DEC = +35$^{\circ}$ 38\,$'$ (J2000.0). Optical spectroscopy was 
performed using the ISIS spectrograph on the William Herschel Telescope 
(WHT)\footnotemark \footnotetext{The WHT is operated by the Isaac Newton Group 
of Telescopes (ING), at the Roque de Los Muchachos Observatory in La Palma, 
Spain.}
and the HET\footnotemark 
Marcario Low 
Resolution Spectrograph (Hill et al.\ 1998). \footnotetext{The Marcario Low Resolution Spectrograph is a joint project of 
the Hobby - Eberly Telescope partnership and the Instituto de Astronomia
de la Universidad Nacional Autonoma de Mexico. The Hobby - Eberly
Telescope is operated by McDonald Observatory on behalf of
The University of Texas at Austin, the Pennsylvania State
University, Stanford University, Ludwig-Maximilians-Universitaet
Muenchen, and Georg-August-Universitaet Goettingen.}
We have reliable spectroscopic 
redshifts for 81\% of the sample, typically from narrow emission lines, with 
median redshift $z_{\rm med}$ = 0.968. For those objects with no spectroscopic 
redshift, 
typically because their optical spectra were completely 
blank, estimates were made based on the $K-z$ relation from Willott et 
al.\ (2003); the validity of this procedure is discussed in Sec. 3. With the 
addition of these estimated redshifts the median redshift for TOOT00 rises to 
1.287; this is a slightly higher value than those found for the 6CE and 7CRS 
samples for both of which $z_{\rm med} \sim$ 1.1 (Rawlings, Eales \& Lacy 
2001 and Willott et al.\ 2002 respectively). Most of these spectra were taken 
at the WHT in August 2000, typically using the technique of `blind' 
spectroscopy (Rawlings, 
Eales \& Warren 1990) in which the radio data were used to determine the 
position and orientation of the spectroscopic slit. The near-infrared images 
were typically taken after spectroscopy and were used to estimate the 
redshifts of the TOOT00 radio sources without spectroscopic redshifts.

\begin{figure}
\resizebox{\hsize}{!}
{\includegraphics[angle = 90]{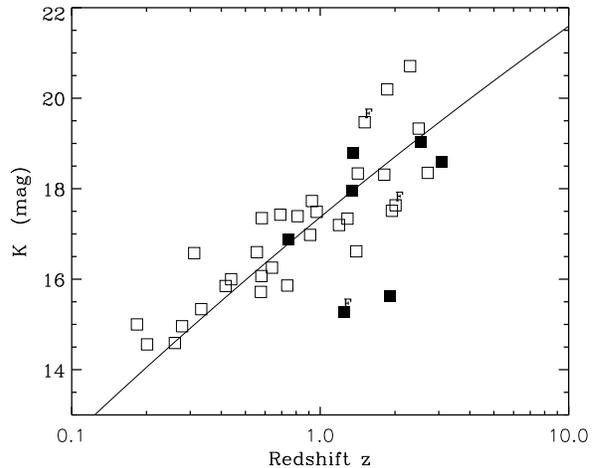}}
\caption{A plot of the 4-arcsec diameter $K$ magnitude versus redshift $z$ 
graph for the 38 TOOT00 radio sources with spectroscopic redshifts. The solid 
line taken from Willott et al.\ 
(2003): $K = 17.37 + 4.53(\log_{10} z) - 0.31(\log_{10} z)^2$; for objects 
with no spectroscopic $z$, a photometric $z$ was estimated from this $K-z$ 
relation. Filled symbols represent quasars and empty symbols galaxies. 
Flat-spectrum radio sources are marked with F.}
\label{k-z}
\end{figure}

For all the TOOT00 objects we have optical spectra from either the WHT or the 
HET, radio maps from the VLA (A and B configurations) and near-infrared data 
from either UKIRT or the Oxford-Dartmouth Thirty Degree (ODT) Survey 
(MacDonald et al.\ 2004). For some of them we obtained multi-colour optical 
images from the ODT survey $K$-band magnitudes are measured from near-infrared 
images for 3, 4, 5, 8 and 9 arcsec diameter apertures, but the 4-arcsec 
aperture value is used in the $K-z$ diagram of Fig.~\ref{k-z}. 

For the TOOT00 radio sources we adopt the radio structure classification from 
Owen \& Laing (1989). We use our VLA data to classify objects as Classical 
Double (CD), Twin Jet (TJ) and Fat Double (FD) radio sources: CDs correspond 
to FRIIs, TJs to FRIs and FDs to FRI/II division objects. The radio spectral 
indices $\alpha_{151{\rm MHz}}^{1.4{\rm GHz}}$ are calculated using the 1.4 
GHz (NRAO/VLA Sky Survey - NVSS) and 151 MHz (7C) flux densities for the TOOT00 sources. The radio 
luminosity at rest-frame 151 MHz $L_{151{\rm MHz}}$ is calculated using the 
flux densities of the TOOT00 objects at 151 MHz, their redshifts and their 
values of $\alpha_{151{\rm MHz}}^{1.4{\rm GHz}}$.

\begin{figure}
\resizebox{\hsize}{!}
{\includegraphics[angle = 90]{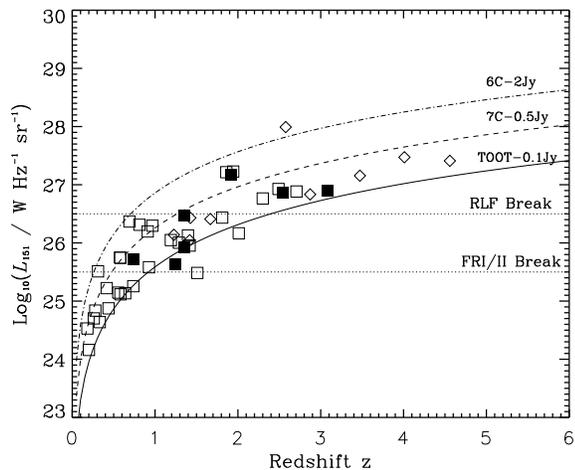}}
\caption{The radio luminosity, $L_{151{\rm MHz}}$, versus redshift $z$ diagram 
for TOOT00. We use square symbols for TOOT00 sources with spectroscopic $z$, 
diamonds for the ones with estimated photometric $z$ from the $K-z$ relation 
(Willott et al.\ 2003), filled symbols for quasars and empty ones for 
galaxies. The over-plotted lines give the flux density limits at 151 MHz of 
the 6CE, 7CRS and TOOT00 samples, which are 2 Jy (Rawlings, Eales \& Lacy 
2001), 500 mJy (Willott et al.\ 2002) and 100 mJy (Hill \& Rawlings 2003) 
respectively.}
\label{l-z}
\end{figure}

\section{Scientific results}

The $K-z$ diagram for the TOOT00 objects with spectroscopic redshifts are 
plotted in Fig.~\ref{k-z}. There is a good agreement with the $K-z$ relation 
from Willott et 
al.\ (2003) that fits the 3CRR, 6CE and 7CRS data. This gives us 
confidence that we can estimate redshifts for the remaining TOOT00 radio 
sources. Furthermore, Willott et al.\ (2003) find that the 
best-fit $K-z$ relation is similar to that expected from a passively evolving 
elliptical galaxy (in a $\Lambda$CDM cosmology) which formed at high redshift 
($\grtsim$ 3). 
From this we conclude that the TOOT00 galaxies, like 6CE and 7CRS galaxies, 
correspond to $\sim$ 3 $L^{*}$ elliptical galaxies.
There are two quasars at the bottom right of the plot which deviate 
significantly from the $K-z$ relation: the $K$ magnitudes of quasars can be 
dominated by continuum emission from the nucleus rather than from a stellar 
population, 
giving a brighter $K$ magnitude. We have ignored any contamination due to 
emission lines since they are unlikely to contribute significantly given the 
known correlation below radio luminosity and emission line strength (e.g. 
Grimes et al.\ 2004), and the low flux density limit of the TOOT.

For all 47 TOOT00 objects we plot the 151-MHz radio luminosity 
$L_{151{\rm MHz}}$  versus redshift diagram (Fig.~\ref{l-z}). The 
over-plotted lines give the flux density limits at 151 MHz of the 6CE, 
7CRS and TOOT samples.

\begin{figure}\resizebox{\hsize}{!}
{\includegraphics[angle = 90]{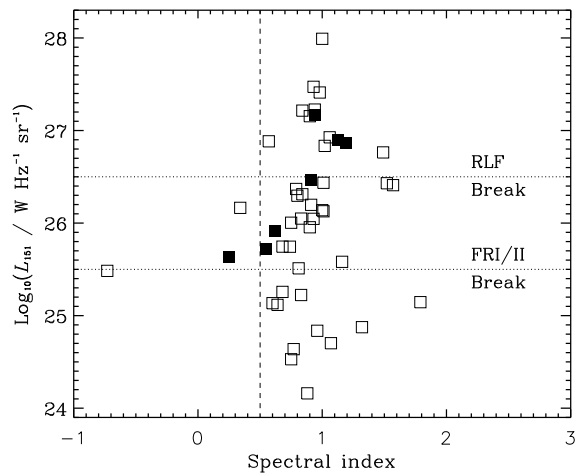}}
\caption{Radio luminosity versus spectral index 
$\alpha_{151{\rm MHz}}^{1.4{\rm GHz}}$ for the TOOT00 radio sources. Filled 
symbols are used for quasars and empty for galaxies. The lower line gives the 
FRI/FRII division at $\log_{10}( L_{151{\rm MHz}})$ = 25.5. 
The upper line is the RLF break at 
$\log_{10}( L_{151{\rm MHz}})$ = 26.5. 
The vertical line gives the flat-steep spectrum division 
$\alpha_{151{\rm MHz}}^{1.4{\rm GHz}}$ = 0.5.}
\label{l-a}
\end{figure}

In Fig.~\ref{l-a} we plot the radio luminosity against the spectral index. 
We take the division between flat and steep spectrum radio 
sources to be at a spectral index 
$\alpha_{151{\rm MHz}}^{1.4{\rm GHz}}$ = 0.5. Then 3 of the 47 TOOT00 radio 
sources are flat spectrum objects\footnotemark. The median spectral index for 
the TOOT00 radio sources $\alpha_{151{\rm MHz}}^{1.4{\rm GHz}}$ = 0.90, which 
tells us that the TOOT00 sources are typically steep-spectrum sources. 
Above the RLF break the quasar fraction, which is defined as the ratio of 
objects with broad emission lines to the total number of them 
(Willott et al.\ 2000), is $f_{\rm q} \sim $ 0.3. 
The spectra of these quasars (Vardoulaki et al.\ in prep) indicate they are 
all reddened, in the sense that the Ly$\alpha$ emission line is 
narrow, and broad ($\grtsim 2000 {\rm km~s^{-1}}$) bases are seen only on 
lines like CIV 1549, implying the 
existence of an obscured nucleus (although spectropolarimetry would be needed 
to disentangle scattered and transmitted light). 
\footnotetext{Two of the tree 
flat-spectrum objects are galaxies; the other one has broad emission lines 
characteristic of a quasar. Note two other quasars lie close to the 
$\alpha_{151{\rm MHz}}^{1.4{\rm GHz}}$ division in Fig.~\ref{l-a}.}
Below the RLF break $f_{\rm q} \rightarrow $ 0 if quasars close to either the 
RLF boundary or the flat/steep spectrum division are excluded. 
Bearing in mind the statistical uncertainty our results are in agreement with 
Willott et al.\ (2000), who found that above the break $f_{\rm q} \sim $ 0.4 
and below the break $f_{\rm q} \sim$ 0. 

\begin{figure}\resizebox{\hsize}{!}
{\includegraphics[angle = -90]{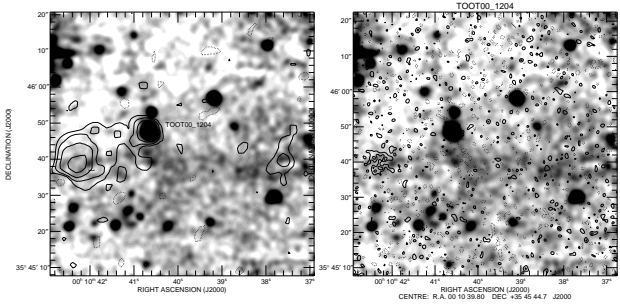}}
\caption{Example of a `fat double' TOOT00 radio galaxy TOOT00\_1204 at 
$z$ = 0.6935, where the B-Array 
(left) and A-array (right) VLA radio 
contours are overplotted on the near-infrared image from the UKIRT.
}
\label{ir-rad}
\end{figure}

In our sample, CDs and FDs typically 
look similar in the B-Array radio maps from the VLA 
(Vardoulaki et al.\ in prep) but the FDs show no compact hotspots in the 
high-resolution maps and most of the integrated flux is clearly in diffuse 
structures (e.g. Fig.~\ref{ir-rad}).

\begin{figure}
\resizebox{\hsize}{!}
{\includegraphics[angle = 90]{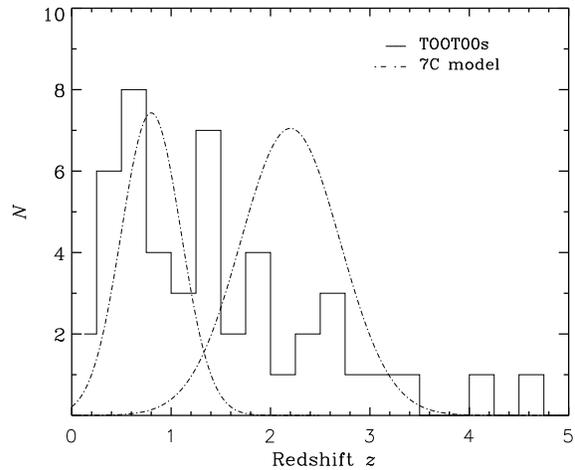}}
\caption{Histogram of the redshift distribution for the 47 TOOT00 radio 
sources with median $z$ = 1.287 (81\% spectroscopic and 19\% photometric 
redshifts). The Willott et al. (2001) model prediction 
for TOOT is approximated by the two Gaussians, which are based on 
model C (for an $\Omega_{\rm M}=0$ Cosmology) of Willott et al.\ (2001) 
normalised such that it reproduces 47 objects in the TOOT00 region.
The Willott et al. model predicted a larger fraction of objects above the RLF 
break and hence at higher redshift in TOOT00.}
\label{hist-z}
\end{figure}

The histogram of the number of TOOT00 objects versus redshift is shown in 
Fig.~\ref{hist-z}. The two TOOT00 objects above $z$ = 4 
have their redshifts estimated from the $K-z$ relation of Willott et al.\ 
(2003) and given the scatter in that relation may not, in reality, be at such 
high redshifts. The overplotted Gaussians approximate the result of the 
Willott et 
al.\ (2001) prediction for TOOT. The prediction considers radio sources in 
two sub-populations, one dominant below the RLF break and one above the 
break. This prediction 
has two maxima, at $z \sim$ 0.8 and 2.2, but predicts more sources in the 
higher redshift population than is observed. In crude terms it appears that 
there is slightly less ICM heating at high redshift due to radio sources 
than predicted by Rawlings (2000) using extrapolations of the RLF similar to 
those presented in Willott et al.\ (2001). We defer full discussion of this, 
as well as subtleties like spatial clustering yielding narrow features in 
N(z), to a future paper.

\section{Conclusions}
The TOOT00 region defines the basic properties of radio sources around the 
RLF break at redshift $\sim$ 1-3. In order to improve the statistics, there is 
a $\sim$10-times larger sample than the TOOT00 region in hand (Hill \& 
Rawlings 2003). In TOOT00 there is a clear correlation between $K$ magnitude 
and redshift $z$, and a good agreement with the $K-z$ relation from Willott et 
al.\ (2003). We are confident of using this relation to estimate rough 
redshifts for the TOOT00 objects with no secure spectroscopic redshifts. The 
host galaxies of the TOOT00 radio sources are similar to 
those at 6CE and 7CRS sources, i.e. $\sim$ 3 $L^{*}$ passively evolved 
elliptical galaxies that formed at high redshift. The 
TOOT00 radio sources are divided into two populations at the RLF break (see 
also Grimes et al.\ 2004): population-2 lies above the RLF break with 
$f_{\rm q} \sim$ 0.3 that basically consists of quasars (sometimes lightly 
reddened) and high-excitation narrow-line radio galaxies; population-1 lies 
below the RLF break and consists of few quasars ($f_{\rm q} \rightarrow$ 0) 
being dominated by low-excitation radio galaxies.

The big remaining question is whether the population-1 
objects have any hidden quasar activity. We know that all TOOT objects have 
central engines producing jets but it is an open question whether jets of 
low power require significant accretion and hence optical quasar activity. 
In order to investigate this we have to study population-1 radio sources with 
Spitzer. Probably the most effective way of doing this is to study 
radio source samples in regions such as the Spitzer First 
Look Region (Condon et al.\ 2003) rather than target regions like TOOT with 
Spitzer.

\acknowledgements
EV and SR would like to thank the organisers of the conference: Montse, Rosa, 
Enrique and Jose Luis for their kind hospitality. EV would also like to thank 
the Ministerio de Educaci\'on y Ciencia for financial support for this 
conference. SR is grateful to the UK PPARC for a Senior Research Fellowship. 
The work of SC was performed in part under the auspices of USDOE by UCLLNL 
under contract W-7405-Eng-48.

\end{document}